\documentstyle[amssymb,preprint,aps]{revtex}

\begin{document}
\title{Hybrid Inflation with Quasi-canonical Supergravity}
\author{C. Panagiotakopoulos}
\address{Physics Division\\
School of Technology\\
University of Thessaloniki\\
Thessaloniki, Greece}
\maketitle

\begin{abstract}
We construct a hybrid inflationary model associated with the superheavy
scale $M_X \simeq 2\times 10^{16}$ $GeV$ of the minimal supersymmetric
standard model which is based on the simplest superpotential for symmetry
breaking and in which the inflaton potential along the inflationary
trajectory is essentially provided by quasi-canonical supergravity. The
resulting spectrum of adiabatic density perturbations is blue and the
duration of inflation sufficient but rather limited.
\end{abstract}

\newpage

Linde's hybrid inflationary scenario \cite{linde94} is certainly superior
over all its predecessors \cite{linde90} because it does not suffer from any
serious generic naturalness problems but also because it succeeded, after a
long time, in reconnecting inflation with phase transitions in grand unified
theories (GUTs). This new inflationary model looks as a hybrid of chaotic
inflation with a quadratic potential for the gauge singlet inflaton and the
usual theory of spontaneous symmetry breaking involving a possibly gauge
non-singlet field. During inflation the non-inflaton field finds itself
trapped in a false vacuum state and the universe expands quasi-exponentially
dominated by the almost constant false vacuum energy density. Inflation ends
with a very rapid phase transition when the non-inflaton field rolls to its
true vacuum state (``waterfall'').

Although the original hybrid model is non-supersymmetric it is so readily
adaptable to supersymmetry (SUSY) that one would have easily thought that it
was invented with SUSY in mind. The simplest and most commonly used
superpotential for symmetry breaking and an inflaton mass of the order of $%
1\ TeV$, the SUSY breaking scale, gives rise to Linde's hybrid model with an
intermediate scale $(\sim 10^{11}-10^{12}\>$ $GeV)$ of symmetry breaking 
\cite{cope}. Moreover, the possibility of imposing R-symmetries in SUSY
models in order to naturally forbid large self-couplings of the inflaton 
\cite{dvali} should be regarded as an additional motivation for
supersymmetry. In the context of supersymmetry it would certainly be
desirable to associate hybrid inflation with the superheavy symmetry
breaking scale $M_{X}\simeq 2\times 10^{16}\>$ $GeV$ which is consistent
with the unification of the gauge coupling constants of the minimal
supersymmetric standard model (MSSM). However, the electroweak mass of the
inflaton provided by SUSY breaking proved too weak to account for the
correct value of the observed temperature fluctuations $\frac{\Delta T}{T}$
in the cosmic background radiation and soon the need for an appropriate
inflaton potential became apparent. The first attempt in this direction was
to employ radiative corrections \cite{dvali} in the context of the simplest
superpotential. This scenario, however, turned out to lead to scales smaller
than the MSSM scale. Afterwards, a variation of the simplest model involving
a non-renormalizable superpotential \cite{laz} was successful in obtaining
the MSSM value of the scale.

Replacing global by local supersymmetry is a highly non-trivial extention
because supergravity makes the potential very steep and typically forbids
inflation through the generation of an inflaton mass larger than the Hubble
constant $H$. For the simplest superpotential, which during inflation could
be regarded as consisting of just a linear term in the inflaton superfield,
the disastrous generation of an inflaton mass-squared term is avoided
provided the canonical form of the K$\ddot{a}$hler potential of $N=1$
supergravity is employed \cite{cope}. However, even in this case
supergravity is expected to affect global SUSY inflationary scenarios,
especially if during inflation the inflaton takes values close to the
supergravity scale $M_{P}/\sqrt{8\pi }\simeq 2.4355\times
10^{18}\>GeV\>\>(M_{P}\simeq 1.221\times 10^{19}\>GeV$ is the Planck mass).

In order to illustrate in the clearest possible way the effects of
supergravity on hybrid inflationary models we investigated \cite{pan} the
possibility that during inflation the inflaton potential is provided
entirely by the terms generated when global supersymmetry is replaced by
canonical supergravity and therefore canonical supergravity is primarily
responsible for the generation of the inflationary density perturbations.
Our supergravity dominated inflationary scenario turned out to have rather
interesting and distinctive properties. Inflation has a very limited but
still sufficient duration and the spectral index of the adiabatic density
perturbations is considerably larger than unity (blue primordial spectra\cite
{mol}) and strongly varying. Although we succeeded in constructing a hybrid
inflationary model associated with the superheavy scale of SUSY GUTs the
MSSM value of the scale could be obtained naturally only in the case of a
model involving two non-inflaton fields. In the context of the simplest
superpotential the MSSM scale was obtained with a choice of very weak
coupling. Larger scales could be obtained more naturally.

The effects of canonical supergravity on the model based on the simplest
superpotential were also investigated in ref. \cite{linde97} in the context
of a combined scenario involving large radiative corrections as well.
Although the combined scenario employs natural values of the parameters the
vacuum expectation value (vev) of the non-inflaton field is typically
smaller than the MSSM scale. To obtain the MSSM value of the scale one has
to make a choice of parameters leading to an unacceptably large spectral
index.

Our purpose in the present paper is to extend our discussion of the
supergravity dominated hybrid inflationary scenario \cite{pan} to the case
of a K$\ddot{a}$hler potential which deviates from the minimal one assuming,
of course, that inflation is still allowed. We confine ourselves to the
simplest superpotential for symmetry breaking and we parametrize the
deviation from canonical supergravity with the size of the generated
inflaton mass-squared term. By allowing deviations from canonical
supergravity, apart from alleviating the fine-tuning that the minimal choice
of the K$\ddot{a}$hler potential necessarily entails, we succeed in
obtaining the MSSM value for the symmetry breaking scale in the context of
the simplest superpotential and for various natural values of the
parameters. The resulting scenario gives rise again to a limited number of
e-foldings and to blue primordial spectra, although there is a tendency for
smaller values of the spectral index. The effects of radiative corrections
are taken into account as well.

We consider a SUSY GUT based on a (semi-simple) gauge group $G$ of rank $%
\geq 5$. $G$ breaks spontaneously directly to the standard model (SM) gauge
group $G_{S}$ $\equiv SU(3)_{c}\times SU(2)_{L}\times U(1)_{Y}$ at a scale $%
M_{X}\sim 10^{16}\>GeV$. The symmetry breaking of $G$ to $G_{S}$ is obtained
through a superpotential which includes the terms 
\begin{equation}
W=S(-\mu ^{2}+\lambda \Phi \bar{\Phi}).
\end{equation}
Here $\Phi ,\bar{\Phi}$ is a conjugate pair of left-handed SM singlet
superfields which belong to non-trivial representations of $G$ and reduce
its rank by their vevs, $S$ is a gauge singlet left-handed superfield, $\mu $
is a superheavy mass scale related to $M_{X}$ and $\lambda $ a real and
positive coupling constant. The superpotential terms in eq. (1) are the
dominant couplings involving the superfields $S$, $\Phi $, $\bar{\Phi}$
which are consistent with a continuous R-symmetry under which $W\to
e^{i\gamma }W$, $S\to e^{i\gamma }S$, $\Phi \to \Phi $ and $\bar{\Phi}\to 
\bar{\Phi}$. Moreover, we assume that the presence of other $SM$ singlets in
the theory does not affect the superpotential in eq. (1). The potential
obtained from $W$, in the supersymmetric limit, is 
\begin{equation}
V=\mid -\mu ^{2}+\lambda \Phi \bar{\Phi}\mid ^{2}+\mid \lambda S\mid
^{2}(\mid \Phi \mid ^{2}+\mid \bar{\Phi}\mid ^{2})+D-terms,
\end{equation}
where the scalar components of the superfields are denoted by the same
symbols as the corresponding superfields. The SUSY vacuum 
\begin{equation}
<S>=0,<\Phi ><\bar{\Phi}>=\mu ^{2}/\lambda ,\>\>\mid <\Phi >\mid =\mid <\bar{%
\Phi}>\mid
\end{equation}
lies on the D-flat direction $\Phi =\bar{\Phi}^{*}$. By appropriate gauge
and R-trasformations on this D-flat direction we can bring the complex $S$, $%
\Phi $, $\bar{\Phi}$ fields on the real axis, i.e. $S\equiv \frac{1}{\sqrt{2}%
}\sigma $, $\Phi =\bar{\Phi}\equiv \frac{1}{2}\phi $, where $\sigma $ and $%
\phi $ are real scalar fields. The potential in eq. (2) then becomes 
\begin{equation}
V(\phi ,\sigma )=(-\mu ^{2}+\frac{1}{4}\lambda \phi ^{2})^{2}+\frac{1}{4}%
\lambda ^{2}\sigma ^{2}\phi ^{2}
\end{equation}
and the supersymmetric vacuum corresponds to $\mid <\frac{\phi }{2}>\mid =%
\frac{\mu }{\sqrt{\lambda }}=\frac{M_{X}}{g}$ and $<\sigma >=0$, where $%
M_{X} $ is the mass acquired by the gauge bosons and $g$ is the gauge
coupling constant. For any fixed value of $\sigma >\sigma _{c}$, where $%
\sigma _{c}=\sqrt{2}\mu /\sqrt{\lambda }=\sqrt{2}\mid <\frac{\phi }{2}>\mid $%
, $V$ as a function of $\phi $ has a minimum lying at $\phi =0$. The value
of $V$ at this minimum for every value of $\sigma >\sigma _{c}$ is $\mu ^{4}$%
.

Adding to $V$ a mass-squared term for $\sigma $ we essentially obtain
Linde's potential. When $\sigma >\sigma _{c}$ the universe is dominated by
the false vacuum energy density $\mu ^{4}$ and expands quasi-exponentially.
When $\sigma $ falls below $\sigma _{c}$ the mass-squared term of $\phi $
becomes negative, the false vacuum state at $\phi =0$ becomes unstable and $%
\phi $ rolls rapidly to its true vacuum thereby terminating inflation.

Let us now replace global supersymmetry by $N=1$ canonical supergravity.
From now on we will use the units in which $\frac{M_{P}}{\sqrt{8\pi }}=1$.
Then, the potential $V(\phi ,\sigma )$ becomes 
\begin{equation}
V(\phi ,\sigma )=[(-\mu ^{2}+\frac{1}{4}\lambda \phi ^{2})^{2}(1-\frac{%
\sigma ^{2}}{2}+\frac{\sigma ^{4}}{4})+\frac{1}{4}\lambda ^{2}\sigma
^{2}\phi ^{2}(1-\frac{\mu ^{2}}{\lambda }+\frac{1}{4}\phi ^{2})^{2}]%
\displaystyle{e^{\frac{1}{2}(\sigma ^{2}+\phi ^{2})}}.
\end{equation}
$V$ still has a minimum with $V=0$ at $\mid \frac{\phi }{2}\mid =\frac{\mu }{%
\sqrt{\lambda }}$ and $\sigma =0$ and a critical value $\sigma _{c}$ of $%
\sigma $ which remains essentially unaltered. The important difference lies
in the expression of $V(\sigma )$ for $\sigma >\sigma _{c}$ and $\phi =0$ 
\begin{equation}
V(\sigma )=\mu ^{4}(1-\frac{\sigma ^{2}}{2}+\frac{\sigma ^{4}}{4})%
\displaystyle{e^{\frac{\sigma ^{2}}{2}}},
\end{equation}
which now is $\sigma $-dependent. Obviously the inflaton potential $V(\sigma
)$ during inflation is obtainable from the simple linear superpotential 
\begin{equation}
W=-\mu ^{2}S,
\end{equation}
with the choice 
\begin{equation}
K=\mid S\mid ^{2}
\end{equation}
for the K$\ddot{a}$hler potential. Expanding $V(\sigma )$ in powers of $%
\sigma ^{2}$ and keeping the first non-constant term only we obtain 
\begin{equation}
V(\sigma )\simeq \mu ^{4}+\frac{1}{8}\mu ^{4}\sigma ^{4}\qquad (\sigma
^{2}<<1).
\end{equation}
We see that no mass-squared term for $\sigma $ is generated \cite{cope}.

Allowing deviations from the canonical form of the K$\ddot{a}$hler potential 
$K$ of eq. (8) which respect the R-symmetry we are led to a K$\ddot{a}$hler
potential 
\begin{equation}
K=\mid S\mid ^{2}-\frac{\beta }{4}\mid S\mid ^{4}+\ldots
\end{equation}
By an appropriate choice of the omitted terms in the expansion of $K$ in eq.
(10) we can arrange for a potential whose expansion in powers of $\sigma
^{2} $ (keeping the first two non-constant terms only) is

\begin{equation}
V(\sigma )\simeq \mu ^{4}+\frac{1}{2}\beta \mu ^{4}\sigma ^{2}+\frac{1}{8}%
\mu ^{4}\sigma ^{4}\qquad (\sigma ^{2}<<1).
\end{equation}
The model now resembles the original hybrid inflationary model with a
quadratic $\frac{1}{2}m^{2}\sigma ^{2}$ term and an additional quartic $%
\frac{1}{4}\kappa \sigma ^{4}$ term, where $m^{2}=\beta \mu ^{4}$ and $%
\kappa =\frac{1}{2}\mu ^{4}.$ The derivative of $V(\sigma )$ with respect to 
$\sigma $ is

\begin{equation}
V^{\prime }(\sigma )\simeq \frac{\mu ^{4}}{2\sigma }(2\beta \sigma
^{2}+\sigma ^{4})\qquad (\sigma ^{2}<<1).
\end{equation}

In the following we are going to study inflation in the context of the
simple model of eq. (1) with an almost-canonical K$\ddot{a}$hler potential $%
K $ identifying the properly normalized inflaton field with $\sigma $ (thus
neglecting the effect of the non-canonical kinetic term) and approximating
the inflaton potential with the constant term only

\begin{equation}
V(\sigma )\simeq \mu ^{4}\text{.}
\end{equation}
For the derivative $V^{\prime }(\sigma )$ of $V(\sigma )$ we are going to
use the expression

\begin{equation}
V^{\prime }(\sigma )\simeq \frac{\mu ^{4}}{2\sigma }\left[ \left( \frac{%
\lambda }{2\pi }\right) ^{2}+2\beta \sigma ^{2}+\sigma ^{4}\right] .
\end{equation}
The first term in the above expression is the contribution of radiative
corrections \cite{dvali} along the inflationary trajectory. We assume that $%
\frac{1}{3}>\beta >\frac{\lambda }{2\pi }$. The inequality $\beta <$ $\frac{1%
}{3}$ is necessary in order that $\frac{m}{H}=\sqrt{3\beta }$ $<1$ ($H=\frac{%
\mu ^{2}}{\sqrt{3}}$ is the Hubble constant during inflation). The
assumption $\beta >$ $\frac{\lambda }{2\pi }$ is made since we are primarily
interested in a scenario in which the deviation from canonical supergravity
is as large as possible and the effect of radiative corrections as small as
possible. The coupling $\lambda $ is determined by requiring that the vev of
the non-inflaton field takes the MSSM value $\mid <\frac{\phi }{2}>\mid =%
\frac{M_{X}}{g}$ $\simeq 0.011731$ ($M_{X}\simeq 2\times 10^{16}$ $%
GeV,g\simeq 0.7$):

\begin{equation}
\lambda =\left( g\frac{\mu }{M_{X}}\right) ^{2}\text{.}
\end{equation}
The critical value $\sigma _{c}$ is fixed by the relation

\begin{equation}
\sigma _{c}=\sqrt{2}\frac{M_{X}}{g}\simeq 0.01659
\end{equation}
(or $\sigma _{c}\simeq 4.0406\times 10^{16}\;GeV$) which holds in our simple
model. Inflation ends through the ``waterfall'' mechanism at $\sigma _{c}$
provided $\sigma _{c}^{2}\gtrsim \frac{\lambda ^{2}}{8\pi ^{2}}$ , i.e. $%
\lambda \lesssim $ $0.147$ or equivalently $\frac{\mu ^{{}}}{M_{X}^{{}}}%
\lesssim 0.548$.

Assuming, as it turns out to be the case, that $(\frac{\Delta T}{T}%
)_{T}^{2}/(\frac{\Delta T}{T})_{S}^{2}<<1$, where $(\frac{\Delta T}{T})_{T}$
and $(\frac{\Delta T}{T})_{S}$ are the tensor and scalar components of the
quadrupole anisotropy $\frac{\Delta T}{T}$ respectively, we identify $\frac{%
\Delta T}{T}$ with $(\frac{\Delta T}{T})_{S}$ and obtain \cite{lid}

\begin{equation}
\frac{\Delta T}{T}\simeq \frac{1}{4\pi \sqrt{45}}(\frac{V^{3/2}}{V^{\prime }}%
)_{\sigma {_{H}}}=\frac{\mu ^{2}\sigma _{H}}{2\pi \sqrt{45}}\left[ \left( 
\frac{\lambda }{2\pi }\right) ^{2}+2\beta \sigma _{H}^{2}+\sigma
_{H}^{4}\right] ^{-1}.
\end{equation}
Here $\sigma _{H}$ is the value that the inflaton field had when the scale $%
\ell _{H}$, corresponding to the present horizon, crossed outside the
inflationary horizon.

Let us define

\begin{equation}
N(\sigma )\equiv \frac{1}{2\sqrt{\beta ^{2}-(\frac{\lambda }{2\pi })^{2}}}%
\text{ }ln\left[ 1+\frac{2\sqrt{\beta ^{2}-(\frac{\lambda }{2\pi })^{2}}}{%
\sigma ^{2}+\beta -\sqrt{\beta ^{2}-(\frac{\lambda }{2\pi })^{2}}}\right] 
\text{.}
\end{equation}
Then the number of e-foldings $\Delta N(\sigma _{in},\sigma _{f})$ for the
time period that $\sigma $ varies between the values $\sigma _{in}$ and $%
\sigma _{f}\>\ (\sigma _{in}>\sigma _{f})$ is given, in the slow roll
approximation, by 
\begin{equation}
\Delta N(\sigma _{in},\sigma _{f})=-\int_{\sigma _{in}}^{\sigma {_{f}}}\frac{%
V}{V^{\prime }}d\sigma =N(\sigma _{f})-N(\sigma _{in})\text{.}
\end{equation}

Let us denote by $\ell _{H}$ the scale corresponding to our present horizon
and by $\ell _{o}$ another length scale. Also let $\sigma _{o}$ be the value
that the inflaton field had when $\ell _{o}$ crossed outside the
inflationary horizon. We define the average spectral index $n(\ell _{o})$
for scales from $\ell _{o}$ to $\ell _{H}$ as 
\begin{equation}
n(\ell _{o})\equiv 1+2ln[(\frac{\delta \rho }{\rho })_{\ell _{o}}/(\frac{%
\delta \rho }{\rho })_{\ell _{H}}]/ln(\frac{\ell _{H}}{\ell _{o}})=1+2ln[(%
\frac{V^{3/2}}{V^{\prime }})_{\sigma _{o}}/(\frac{V^{3/2}}{V^{\prime }}%
)_{\sigma _{H}}]/\Delta N(\sigma _{H},\sigma _{o}).
\end{equation}
Here $(\delta \rho /\rho )_{\ell }$ is the amplitude of the energy density
fluctuations on the length scale $\ell $ as this scale crosses inside the
postinflationary horizon and $\Delta N(\sigma _{H},\sigma _{o})=N(\sigma
_{o})-N(\sigma _{H})=ln(\ell _{H}/\ell _{o})$.

It should be clear that all quantities characterizing inflation in our
scenario, such as $\sigma _{H}$, the (average) spectral index $n$ and the
number of e-foldings $N_{H}\equiv \Delta N(\sigma _{H},\sigma _{c})$ for the
time period that $\sigma $ varies between $\sigma _{H}$ and $\sigma _{c}$,
depend on just two parameters, namely $\mu $ and $\beta $ or equivalently
the Hubble constant $H=\frac{\mu ^{2}}{\sqrt{3}}$ and the ratio $\frac{m}{H}=%
\sqrt{3\beta \text{ }}$ of the inflaton ``mass'' $m$ to $H$. Therefore for
each value of $\mu $ we can determine $\beta $ by requiring that $N_{H}$
takes the appropriate value. For $0.1\lesssim \frac{\mu }{M_{X}}\lesssim 1$
we choose $N_{H}\simeq 55$.

Table 1 gives the values of $\beta ,$ $\lambda ,$ $\sigma _{H},$ $n\equiv
n(\ell _{1})$ and $n_{COBE}\equiv n(\ell _{2})$, where $\ell _{1}$ $(\ell
_{2})$ is the scale that corresponds to $1$ $Mpc$ $(2000$ $Mpc)$ today, for
different values of $\mu $ together with the chosen value of $N_{H}$
assuming that the present horizon size is $12000$ $Mpc$ and $\frac{\Delta T}{%
T}=6.6\times 10^{-6}$. We see that for a wide range of values of the mass
scale $\mu $ our simple model is able to accomodate the MSSM scale with $%
\beta >$ $\frac{\lambda }{2\pi }$ (and even with $\beta >\lambda $ provided $%
\mu \lesssim 5\times 10^{15}$ $GeV$). Also for $2\times 10^{15}$ $%
GeV\lesssim \mu \lesssim 8\times 10^{15}$ $GeV$ we have $0.021\lesssim \beta
\lesssim 0.031$ and consequently $0.25\lesssim \frac{m}{H}\lesssim 0.31$.
Comparing with the case of canonical supergravity we believe that the
improvement concerning naturalness is quite impressive. The spectrum of
density perturbations is blue, like in the case of canonical supergravity,
with an apparent tendency, however, for lower values of the spectral index.
As a result of this lowering values of $\mu $ almost as high as $9\times
10^{15}$ $GeV$ are now consistent with the COBE data. Moreover, there is
also a tendency for lower values of $\sigma _{H}$ due to the contribution of
the inflaton mass-squared term. This effect becomes more important as $\mu $
decreases.

One can understand the drastic lowering of $\sigma _{H}$ for ``small'' $\mu $
by observing that the quadratic term in the bracket of eq. (17) dominates
for $\frac{\lambda }{2\pi }\frac{1}{\sqrt{2\beta }}\lesssim \sigma
_{H}\lesssim \sqrt{2\beta }$. Also we see that $\sigma _{H}$ decreases with $%
\mu $ provided $\sigma _{H}^{2}\gg \left( \frac{\lambda }{2\pi }\right) ^{2}%
\frac{1}{2\beta }$. Therefore as $\mu $ decreases $\sigma _{H}$ approaches $%
\sqrt{2\beta }$ and the serious contribution of the quadratic term results
in a drastic lowering of $\sigma _{H}$ relative to the canonical
supergravity scenario.

One expects that with increasing $\lambda $ radiative corrections will start
playing an increasingly important role towards the end of inflation. For $%
\sigma _{c}\gtrsim \frac{\lambda }{2\pi }\frac{1}{\sqrt{2\beta }}$, or $\mu
\lesssim 4\times 10^{15}$ $GeV$, the radiative correction term is never
dominant in $V^{\prime }\left( \sigma \right) $ and its effect on inflation
is expected to be rather limited. For $\mu \gtrsim 5\times 10^{15}$ $GeV$ we
see that $\lambda >\beta $ and $\beta $ starts falling slowly with $\lambda $
increasing. The peak of $\beta $ around $\mu \simeq 5\times 10^{15}$ $GeV$
is actually a result of radiative corrections. To understand better their
role in our scenario we repeated all calculations neglecting radiative
corrections altogether. Table 2 contains the results of this investigation
for some values of $\mu $. Comparing the corresponding values listed in
Table 1 and Table 2 we see that the suppression of $\beta $ with increasing $%
\mu $ is clearly due to the contribution of radiative corrections. Apart
from their negative contribution to naturalness radiative corrections have
also a minor effect on the spectral index and a somewhat larger one on $%
\sigma _{H}$. Both these effects, however, are due to the suppression of $%
\beta $ since radiative corrections, which become more important for $\sigma
\lesssim \frac{\lambda }{2\pi }\frac{1}{\sqrt{2\beta }}$, could not directly
affect $\sigma _{H}$ or the spectral index. We conclude that the blue
perturbation spectra and the success of our scenario in obtaining the MSSM
scale in the context of the simplest model should be attributed primarily to
supergravity.

Extrapolating the results listed in Table 1 for $\mu \ll 10^{15}$ $GeV$ we
reach the weak coupling canonical supergravity scenario of ref.\cite{pan} in
the context of the simplest model. An extrapolation for $\mu \gtrsim 10^{16}$
$GeV$ shows how one could obtain the MSSM scale in the context of the
scenario of ref. \cite{linde97} with large radiative corrections and
canonical supergravity. Notice that in both scenarios for each value of $\mu 
$ the only parameter left, namely $\lambda $, is fixed once the exact value
of $N_{H}$ is chosen. Consequently, in order to obtain the MSSM scale with
canonical supergravity and in the context to the simplest model, one is left
only with the choice of $\mu $ and with the very general choice of weak or
strong coupling or equivalently the choice of the scenario.

In the above discussion we only considered a one-parameter deviation of the K%
$\ddot{a}$hler potential from its canonical form. It is understood that
different choices of the omitted terms in eq. (10) could possibly lead to a
further improvement concerning naturalness.

The evolution of the universe after the ``waterfall'' cannot be addressed on
general terms since it depends crucially on the details of the specific
particle physics model incorporating the present inflationary scenario. In
the context of a concrete model one should discuss the numerous difficult
issues that the subsequent evolution involves like the ``reheat'', the
gravitino problem, the generation of the baryon asymmetry, the existence of
suitable dark matter candidates, the formation of topological defects etc. A
detailed discussion of these issues is clearly beyond the scope of the
present paper.

We conclude by summarizing our results. We considered a supergravity hybrid
inflationary scenario in the context of the simplest superpotential giving
rise to symmetry breaking. By allowing deviations from the minimal K$\ddot{a}
$hler potential we succeeded in obtaining the MSSM value of the symmetry
breaking scale for several natural values of the parameters and with an
inflaton ``mass'' only three to four times smaller than the Hubble constant.
The spectrum of adiabatic density perturbations is blue and the duration of
inflation rather limited. We believe that our quasi-canonical supergravity
scenario is the most natural realization of Linde's hybrid inflation in the
context of supersymmetry.

\acknowledgments

This research was supported in part by EU under TMR contract
ERBFMRX-CT96-0090. The author would like to thank G. Lazarides for useful
discussions.

\newpage 
\begin{tabular}{ccccccc}
$\mu /10^{15}GeV$ & $N_{H}$ & $\qquad \beta \qquad $ & $\qquad \lambda
\qquad $ & $\sigma _{H}/10^{17}GeV$ & $\quad n\quad $ & $n_{COBE}$ \\ 
&  &  &  &  &  &  \\ 
1 & 56.12 & 0.00987 & 0.0012 & 0.7149 & 1.022 & 1.022 \\ 
2 & 55.63 & 0.0205 & 0.0049 & 1.3355 & 1.048 & 1.049 \\ 
3 & 55.29 & 0.0273 & 0.0110 & 2.1221 & 1.071 & 1.076 \\ 
4 & 55.16 & 0.0307 & 0.0196 & 3.0422 & 1.094 & 1.104 \\ 
5 & 54.99 & 0.0312 & 0.0306 & 4.0556 & 1.118 & 1.138 \\ 
6 & 54.74 & 0.0294 & 0.0441 & 5.1111 & 1.145 & 1.178 \\ 
7 & 54.68 & 0.0258 & 0.0600 & 6.1704 & 1.173 & 1.224 \\ 
8 & 54.49 & 0.0212 & 0.0784 & 7.1910 & 1.203 & 1.275 \\ 
9 & 54.43 & 0.0158 & 0.0992 & 8.1697 & 1.233 & 1.328
\end{tabular}

\bigskip

\bigskip

\bigskip

Table 1. The values of $N_{H}$, $\beta $, $\lambda $, $\sigma _{H}$, $n$ and 
$n_{COBE}$ as a function of $\mu $.

\bigskip

\bigskip

\bigskip

\bigskip

\qquad

\medskip

\begin{tabular}{cccccc}
$\mu /10^{15}GeV$ & $N_{H}$ & $\qquad \beta \qquad $ & $\sigma
_{H}/10^{17}GeV$ & $\quad n\quad $ & $n_{COBE}$ \\ 
&  &  &  &  &  \\ 
3 & 55.30 & 0.0285 & 2.0687 & 1.073 & 1.077 \\ 
5 & 54.92 & 0.0380 & 3.7165 & 1.122 & 1.140 \\ 
7 & 54.73 & 0.0432 & 5.3622 & 1.173 & 1.216 \\ 
9 & 54.48 & 0.0464 & 6.9042 & 1.223 & 1.301 \\ 
&  &  &  &  & 
\end{tabular}

\medskip

Table 2. The values of $N_{H}$, $\beta $, $\sigma _{H}$, $n$ and $n_{COBE}$
as a function of $\mu $,

ignoring radiative corrections.\qquad

\qquad

\end{document}